\documentclass{aa}  

\usepackage{graphicx}
\usepackage{txfonts}
\begin{document}

   \title{Low-velocity large-scale shocks in the infrared dark cloud G035.39-00.33: bubble-driven cloud-cloud collisions}

      \author{G. Cosentino\inst{1}\thanks{E-mail:giuliana.cosentino@iram.fr},
        I. Jim\'{e}nez-Serra\inst{2},
        R. Liu\inst{3},
        C.-Y. Law\inst{4},
        J. C. Tan\inst{5,6},
        J. D. Henshaw\inst{7},
        A. T. Barnes\inst{8},
        F. Fontani\inst{4,9,10},\\
        P. Caselli\inst{9},
        S. Viti\inst{11}}
    \authorrunning{Cosentino et al.}
    \titlerunning{Cloud-cloud Collision in the IRDC G035.39}
    \institute{Institut de Radioastronomie Millimétrique, 300 Rue de la Piscine, 38400 Saint-Martin-d’Hères, France
    \and Centro de Astrobiolog\'{i}a (CSIC/INTA), Ctra. de Torrej\'on a Ajalvir km 4, Madrid, Spain
    \and National Astronomical Observatories of China, Chinese Academy of Sciences, Beijing, 100012, China
    \and INAF  Osservatorio Astronomico di Arcetri, Largo E. Fermi 5, 50125 Florence, Italy
    \and Department of Space, Earth and Environment, Chalmers University of Technology, SE-412 96 Gothenburg, Sweden
    \and Department of Astronomy, University of Virginia, 530 McCormick Road Charlottesville, 22904-4325 USA
    \and Astrophysics Research Institute, Liverpool John Moores University, 146 Brownlow Hill, Liverpool L3 5RF, UK
    \and European Southern Observatory (ESO), Karl-Schwarzschild-Straße 2, 85748 Garching bei M\"unchen, Germany
    \and Max Planck Institute for Extraterrestrial Physics, Giessenbachstrasse 1, 85748 Garching bei M\"{u}nchen, Germany
    \and Laboratory for the study of the Universe and eXtreme phenomena (LUX), Observatoire de Paris, 5, place Jules Janssen, 92195 Meudon, France
    \and Leiden Observatory, Leiden University, PO Box 9513, 2300 RA Leiden, The Netherlands}

   \date{Received ---; accepted ---}

   \date{Received September 15, 1996; accepted March 16, 1997}

 
  \abstract
   {Low-velocity, large-scale shocks impacting on the interstellar medium have been suggested as efficient mechanisms that shape molecular clouds and trigger star formation within them.}
   {These shocks, both driven by galactic bubbles and/or cloud-cloud collisions, leave specific signatures in the morphology and kinematics of the gas. Observational studies of such signatures are crucial to investigate if and how shocks affect the clouds formation process and trigger their future star formation.}
   {We have analysed the shocked and dense gas tracers SiO(2-1) and H$^{13}$CO$^+$(1-0) emission toward the Infrared Dark Cloud G035.39-00.33, using new, larger-scale maps obtained with the 30 m telescope at the Instituto de Radioastronom\'ia Millim\'etrica.}
   {We find that the dense gas is organised into a northern and a southern filament having different velocities and tilted orientation with respect to each other. The two filaments, seen in H$^{13}$CO$^+$, are spatially separated yet connected by a faint bridge-like feature also seen in a position-velocity diagram extracted across the cloud. This bridge-feature, typical of cloud-cloud collisions, also coincides with a very spectrally narrow SiO-traced gas emission. The northern filament is suggested to be interacting with the nearby supernova remnant G035.6-0.4. Toward the southern filament, we also report the presence of a parsec-scale, spectrally narrow SiO emission likely driven by the interaction between this filament and a nearby expanding shell. The shell is visible in the 1.3 GHz and 610 MHz continuum images and our preliminary analysis suggests it may be the relic of a supernova remnant.}
   {We conclude that the two filaments represent the densest part of two colliding clouds, pushed toward each other by nearby Supernova Remnants. We speculate that this cloud-cloud collision driven by stellar feedback may have assembled the infrared dark cloud. We also evaluate the possibility that star formation may have been triggered within G035.39-00.33  by the cloud-cloud collision.}

   \keywords{ISM: bubbles – ISM: clouds – ISM: molecules – ISM: supernova remnants – ISM: individual objects: G035.39-00.33}

   \maketitle
%

\section{Introduction}
Infrared Dark Clouds (IRDCs) are among the densest \citep[n(H$_2)>$10$^3$ cm$^{-3}$;][]{chevance2023}, coldest \citep[T$<$25 K;][]{pillai2006, rathborne2006} and most-extinct \citep[A$_{\rm{v}}$>100 mag;][]{butlerTan2009,butlerTan2012} regions of the interstellar medium. First seen as dark silhouettes against the mid-infrared Galactic background \citep{egan1998,perault1996}, IRDCs are known to harbour the initial conditions of star and stellar cluster formation at a wide range of masses \citep{kauffmann2010,tan2014,moser2020}. However, the mechanisms that ignite star formation in these objects are still under debate \citep[see][for a review]{chevance2023}. Among different scenarios, parsec-scale low-velocity (10-20 km s$^{-1}$) shocks have been proposed as key ingredients for the star-formation potential of IRDCs \citep{inutsuka2015,inoueFukui2018,Kinoshita2021} that can compress the gas and trigger star formation in clouds. These shocks can develop at the interface of the collisions between pre-existent Giant Molecular Clouds \citep[cloud-cloud collisions;][]{tan2000,tasker2009,dobbs2015,wu2017,wu2020,Horie2024} and/or be driven by the expanding shells of H{\small II} regions and supernova remnants (SNRs) at late stages \citep{fukui2018,cosentino2019,reach2019,RicoVillas2020,Khullar2024}.\\
Observationally, shocks can be identified by means of molecular species whose chemistry is susceptible to their effects. Among these, silicon monoxide (SiO) is a unique tracer of SiO-traced gas usually extremely depleted in quiescent regions \citep[$\chi<$10$^{-12}$;][]{martinpintado1992}, but whose abundance is significantly enhanced by shocks propagating through dense clouds \citep{jimenezserra2005,nath2008,guillet2011}. Indeed, shocks cause sputtering and erosion of dust grains and their icy mantles, releasing Si into the gas phase \citep{caselli1997,schilke1997}. The spatial morphology and line profiles of SiO are expected to be indicative of the spatial extent and velocity of the shock driving the emission \citep{jimenezserra2005,jimenezserra2010,cosentino2020}. For instance, toward molecular outflows, the SiO emission reflects the high velocities and small spatial scales of the shocks, i.e. it is expected to be spatially compact and spectrally broad \citep[e.g.][]{codella2013,liu2020,GuerraVaras2023}. On the other hand, in cloud-cloud collisions, shocks are expected to extend over parsec scales and to have velocities $\sim$10 km s$^{-1}$ \citep{tasker2009,LiuTie2018}. As a consequence, SiO emission toward these regions is expected to be spatially widespread and spectrally narrow \citep[][]{cosentino2018,cosentino2022,kim2023}. Cloud-cloud collisions can also be identified observationally through the detection of the so-called "bridge-like feature" in the low- and high-density gas kinematics. When collisions occur, position-velocity (pv) diagrams along the line of sight are expected to show two main velocity structures, corresponding to the colliding clouds, connected by a fainter bridge emission, i.e. the gas at the interface of the collision. The “bridge feature” has been extensively predicted by simulations \citep[e.g. ][]{Haworth2015a, Haworth2015MNRASb} and observed toward several sources \citep[e.g. ][]{duarteCabral2011,Nakamura2012,Dewangan2018,Tokuda2019,Gong2019,Zeng2020, Ma2022,Kohno2025}. The presence of a bridge feature accompanied by the detection of narrow and widespread SiO-traced gas emission \citep{jimenezserra2010,cosentino2020} and/or bright dense gas emission \citep{priestley2021} is usually considered a more conclusive observational proof of cloud-cloud collision. Shocks driven by galactic bubbles, i.e., H{\small II} regions and late stage SNRs, are also expected to be extended over parsec scales \citep{inutsuka2015}, but their velocities can reach few tens of km s$^{-1}$ \citep{sashida2013,cosentino2019,Reach2024}. Hence, the associated SiO emission may be relatively broader than in cloud-cloud collisions. Shocks driven by galactic bubbles can impact already existent nearby IRDCs \citep{inoueFukui2018} and therefore could explain the presence of non-coeval star populations that are commonly found in molecular clouds \citep[e.g.][]{jerabkova2019}.\\

\noindent
A systematic search of widespread an narrow SiO emission has been reported by \cite{jimenezserra2010} and \cite{cosentino2018,cosentino2020} toward the well-known sample of 10 IRDCs presented by \cite{butlerTan2009,butlerTan2012}. Five out of the ten sources have been reported by the authors to show SiO emission that is widespread over a parsec scale and whose line profiles are narrower than $\sim$3 km s$^{-1}$, i.e., significantly narrower than what observed in molecular outflows. Inspection of these clouds surroundings at multiple IR and radio wavelengths have shown that all the five sources are associated and most likely interacting with extended bubbles in the form of H{\small II} regions and/or SNRs \citep[][; Liu et al. in prep]{cosentino2019,cosentino2020}. 
Within this sample, the IRDC G035.39-00.33, or cloud H following the nomenclature of \cite{butlerTan2009,butlerTan2012}, is a highly filamentary dense cloud whose northern part (thereafter G035.39-N) has been extensively studied \citep[e.g. ][]{jimenezserra2010,NguyenLuong2011,Sokolov2017,LiuTie2018}. It shows a very complex kinematic structure with the molecular gas being distributed in multiple  sub-filaments separated in velocity space by just a few km s$^{-1}$ \citep{henshaw2013,henshaw2014,jimenezserra2014}. CO is highly depleted across the cloud (depletion factors larger than 5), indicating the predominance of a cold chemistry, typical of the initial conditions of star formation \citep{hernandez2011}. As a consequence of the CO depletion, deuterated species such as N$_2$D$^+$ are enhanced across the cloud, with an average D/H ratio 0.04 \citep{barnes2016}. Toward G035.39-N, the IRAM-30m maps of the SiO(2-1) have revealed the presence of both a spectrally broad SiO emission, spatially associated with known embedded young stellar objects within the cloud and a narrow SiO component (linewidth $\sim$1 km s$^{-1}$) widespread across the IRDC \citep{jimenezserra2010}. This narrow and widespread SiO emission was suggested to be the result of a cloud-cloud collision event \citep{jimenezserra2010,bisbas2017}. Follow-up ALMA observations toward G035.39-N seems to indicate that the emission may be instead produced by the ongoing interaction between the cloud and the nearby SNR G035.6-0.4 (Liu et al. in prep.). Indeed, recent JCMT SCUBA2 images at 850 $\mu$m \citep{Shen2024} and MeerKat 1.3 GHz \citep{Goedhart2024} observations have revealed that G035.39 is part of a larger complex, consisting of 5 molecular clouds embedded in an intricate network of expanding bubbles.\\ In this work, we extend the study of the SiO emission toward the south of G035.39 (thereafter G035.39-S). We analyse the spatial distribution and line profile of the SiO emission across G035.39-south and investigate the mechanism through which this emission may be originating. For this, we also investigate the morphology and kinematics of the dense gas tracer H$^{13}$CO$^+$. The paper is organised as follow. In Section~\ref{obs}, we discuss the data acquisition strategy and ancillary data used. In Section~\ref{results}, we present our result. Finally, in Sections~\ref{discussion}, we discuss our findings and draw our conclusions.

\begin{figure*}
    \centering
    \includegraphics[width=\linewidth,trim = 2cm 0cm 3cm 0cm, clip=True]{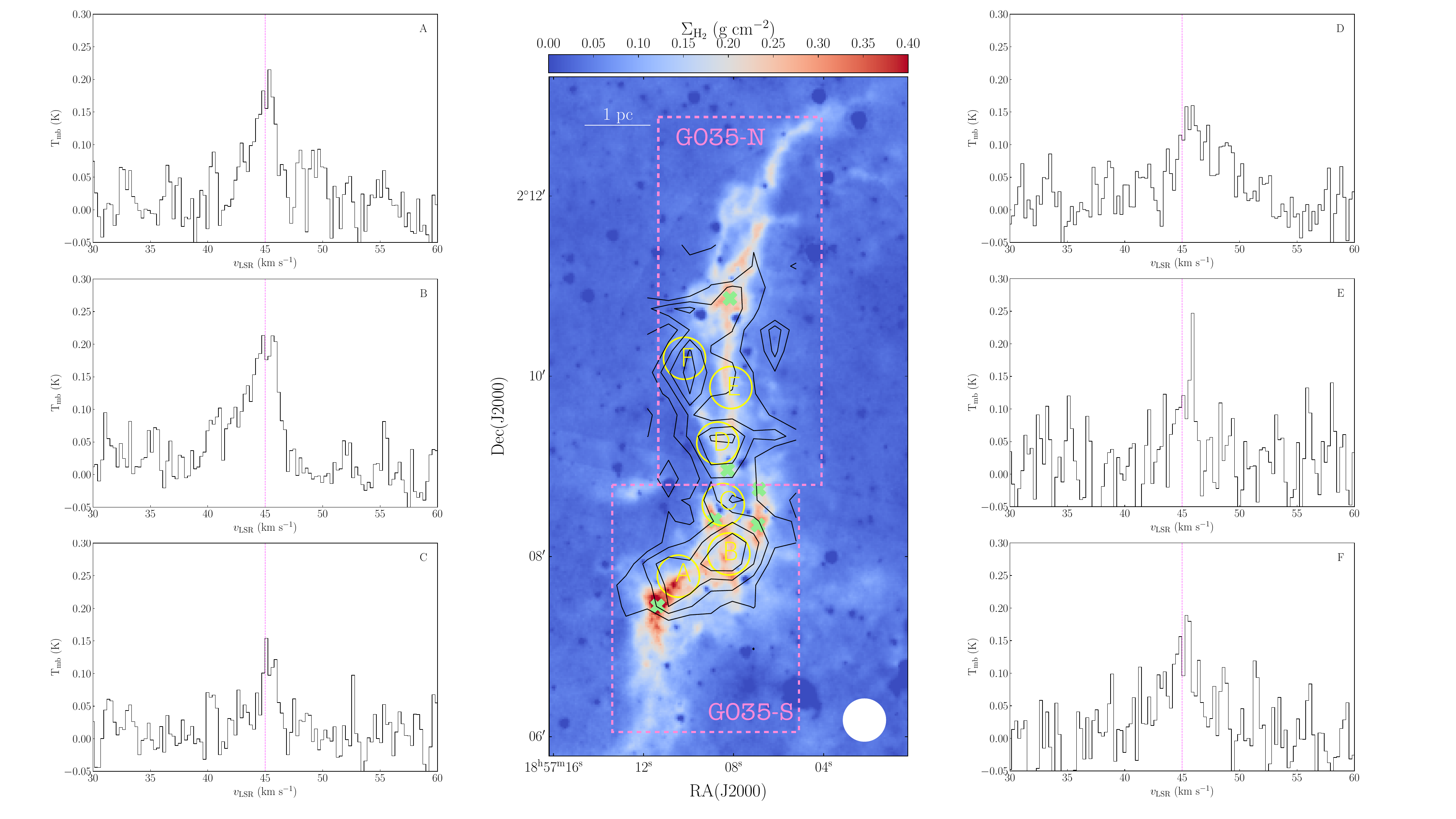}
    \caption{\textit{Middle:} SiO(2-1) integrated intensity map (black contours) obtained toward the velocity range 38-50 km s$^{-1}$. The emission contours, from 3$\sigma$ ($\sigma$=0.07 K km s$^{-1}$) by steps of 3$\sigma$, are superimposed on the mass surface density map (color scale) obtained by \cite{kainulainen2013}. The beam size and spatial scale are indicated in the bottom-right and top-left corners of the map, respectively. The location of the massive cores within the cloud is indicated with light green crosses \citep{rathborne2006,butlerTan2012}. The yellow circles, labelled from A to F, indicates the positions used to extract the SiO spectra. The pink dashed square separate the northern and southern parts of the cloud. \textit{Right and Left:} SiO(2-1) spectra extracted toward the six positions from A to F and using an aperture of 28$^{\prime\prime}$, the size of the beam of the IRAM 30m telescope at the frequency of the SiO(2-1) line. In each panel, the cloud velocity \citep[v$_{\mathrm{LSR}}$=45.5 km s$^{-1}$][]{hernandez2015} is indicated by a vertical dashed magenta line.}
    \label{figSiO}
\end{figure*}

\section{Observations and Data}\label{obs}
\subsection{IRAM-30m Data}
In May 2013, the emission from SiO(2-1) and H$^{13}$CO$^+$(1-0) molecular lines was mapped toward the full extent of G035.39, using the EMIR receiver ($\nu$=86.84696 GHz) on the 30m antenna at the Instituto de Radioastronomia Millimetrica (Pico Veleta, Spain). Observations were performed in on-the-fly (OTF) mode using angular separation in the direction perpendicular to the scanning direction of 6$^{\prime\prime}$. The obtained map has central coordinates RA(J2000)=18$^h$57$^m$08$^s$, Dec(J2000)=2$^d$10$^m$30$^s$ and extent 300$^{\prime\prime}\times$480$^{\prime\prime}$. During observations, an off position at (1830$^{\prime\prime}$, 658$^{\prime\prime}$) with respect to the centre coordinates was used. During observations, the Versatile Spectrometric and Polarimetric Array (VESPA) was set to provide spectral resolution of $\sim$40 kHz, corresponding to a velocity resolution of $\sim$0.14 km s$^{-1}$ at the observed frequencies. Intensities were measured in units of antenna temperature, T$^*_A$ and converted into main beam temperatures, \hbox{T$_{\rm{mb}}$= T$^*_A$ (B$_{eff}$/F$_{eff}$)}, using beam and forward efficiencies of B$_{eff}$=0.81 and F$_{eff}$=0.95, respectively. The final cubes were produced using the {\sc class} and {\sc mapping} software within the {\sc gildas} package and have native resolution of 28$^{\prime\prime}$, pixel size of 14$^{\prime\prime}$ and rms per beam per channel of 55 mK. In order to improve the signal-to-noise ratio, all spectra were smoothed to a velocity $\sim$0.25 km s$^{-1}$, producing a final rms per beam per channel of 40 mK. Finally, a first grade polynomial baseline was subtracted from all spectra, pixel by pixel. We note that the molecular tracer emission obtained toward G035.39-N is consistent with that already presented by \cite{jimenezserra2010}, obtained using the previous generation ABCD receivers on the IRAM-30m.

\subsection{Green Bank Telescope Data}
Observations of the C$^{18}$O(1-0) and $^{13}$CO(1-0) emission towards G035.39 were performed in April 2021 (project code SOF09104), using the 100m antenna at the Green Bank Telescope (GBT, West Virginia, USA). For the observations, the Argus 16-element array was used with a 4$\times$4 configuration, where each element was separated by 30.4$^{\prime\prime}$ in the plane spanned by the elevation and cross-elevation. Observations were performed in a fast mapping method that scanned the sky toward the Galactic Longitude in parallel rows with a width of 5.58$^{\prime\prime}$. The final maps have central coordinates RA(J2000) = 34$^o$44$^{\prime}$04$^{\prime\prime}$, DEC(J2000)= -0$^o$33$^{\prime}$18$^{\prime\prime}$, beam size of $\sim$ 7$^{\prime\prime}$ and a velocity resolution of $\sim$0.19 km s$^{-1}$. The average integrated noise was estimated to be 0.294 K km/s for $\rm ^{13}CO$ and 0.209 K km/s for $\rm C^{18}$O. We note that the C$^{18}$O(1-0) dataset is only minimally used and that a comprehensive analysis of the lower-density gas emission is beyond the scope of this work but will be presented in a forthcoming paper by Law et al. in prep.

\section{Results}\label{results}
\subsection{Low-velocity, large-scale shocks in G035.39}
We first investigate the morphology and line profile of the SiO-traced gas across the IRDC G035.39, with particular focus to the southern part of the cloud. In Figure~\ref{figSiO} we show the SiO(2-1) integrated intensity (38-50 km s$^{-1}$) emission contours (black) as overlaid on the mass surface density map obtained by \cite{kainulainen2013} (central panel, color scale). From Figure~\ref{figSiO}, the cloud shows a very filamentary structure with the northern filament almost vertical (oriented toward the north-south direction) and the southern one tilted by almost 45$^{\circ}$ with respect to it. The SiO emission is widespread across the cloud and follows well its morphology. In particular, the newly mapped southern SiO emission also appears tilted by almost 45$^{\circ}$ with respect to the northern emission, which is instead oriented toward the north-south direction (almost vertical). Toward G035.39-S, the location of the SiO emission peak does not coincide with the positions of known cores within the cloud \citep[cyan crosses;][]{rathborne2006,butlerTan2012} and it is extended across 2.4$^{\prime}$, which corresponds to $\sim$2 pc at the source distance \citep[2.9 kpc;][]{simon2006}. In Figure~\ref{figSiO}, SiO spectra extracted toward a few positions in the north (right panels) and in south (left panels) of the cloud are shown. Positions A, B and C are representative locations across the elongated southern SiO emission. Positions D, E and F coincide with those investigated by \cite{jimenezserra2010}. Toward G035.39-N, both the broad (up to $\sim$20 km s$^{-1}$; spectra D and F) and narrow ($\leq$2 km s$^{-1}$; spectrum E) SiO emission already identified by \cite{jimenezserra2010} are clearly seen. Follow-up ALMA observations of this emission will be discussed in details in the forthcoming paper by Liu et al. (in prep.). Toward G035.39-S, the narrowest SiO emission is observed toward position C, i.e., FWHM$\sim$1.8 km s$^{-1}$, slightly larger than that of the narrow SiO emission reported by \cite{jimenezserra2010}. The brightest emission in G035.39-S is instead observed toward positions A and B. Toward these two positions, the spectra are slightly broader than toward position C ($\sim$3 vs $\sim$2 km s$^{-1}$), but still much narrower than what reported toward the north of the cloud, i.e., positions D and F, where Liu et al. in prep. and \cite{jimenezserra2010} reported linewidths as broad as $\sim$20 km s$^{-1}$. In light of all this, the SiO emission toward G035.39-S can be considered as widespread over $\sim$2 pc and spectrally narrower than what seen toward previously identified molecular outflows within the source. 

\subsection{Dynamical properties of the SiO-traced gas}
\begin{figure*}
    \centering
    \includegraphics[width=\linewidth, trim= 3cm 0.5cm 1.5cm 0cm, clip=True]{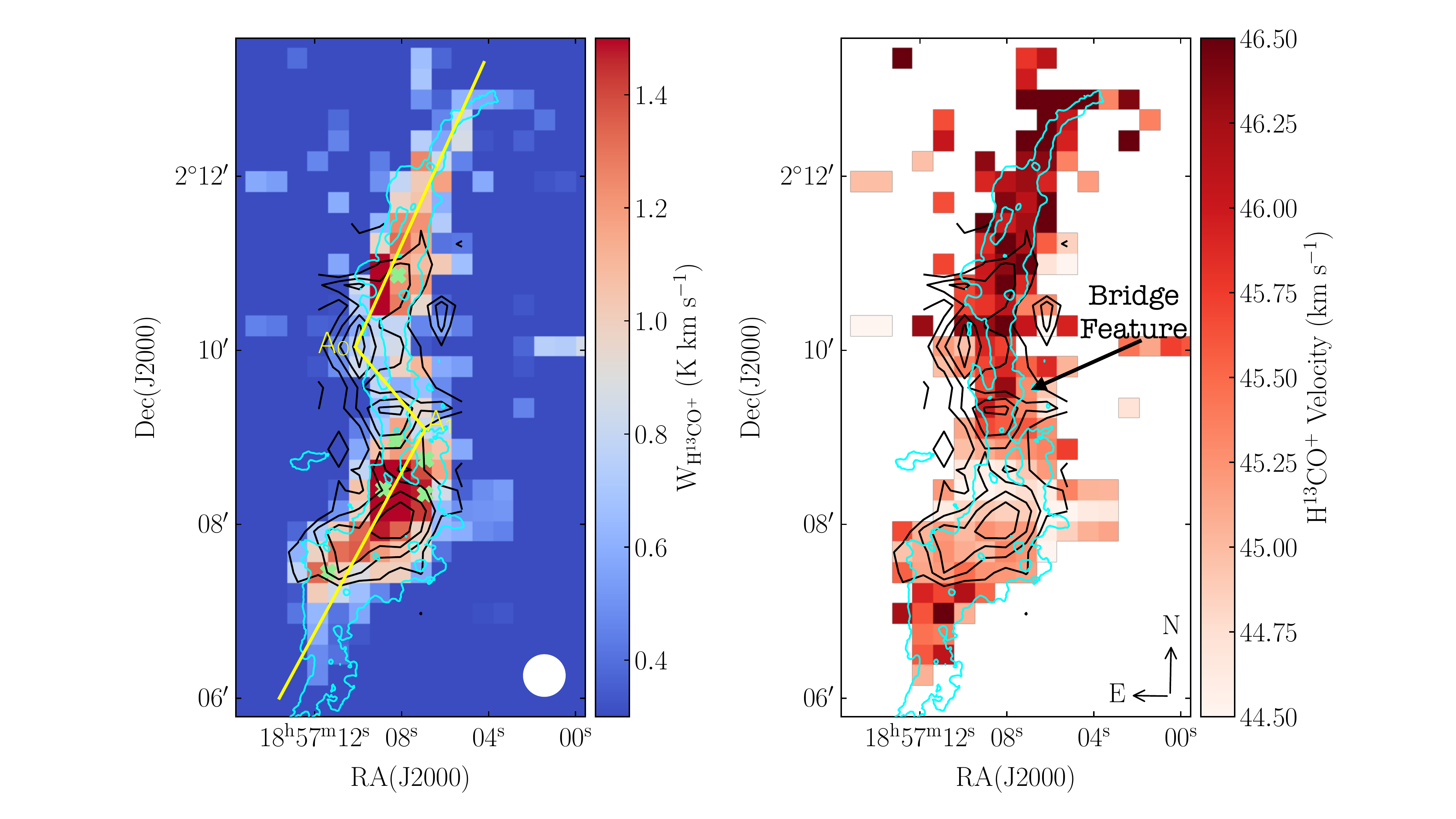}
    \caption{\textit{Left:} H$^{13}$CO$^+$(1-0) integrated intensity map (color scale) obtained toward the velocity range 42-48 km s$^{-1}$. The zero-level of the colorbar  corresponds to 3$\sigma$ ($\sigma$=0.1 K km s$^{-1}$). The yellow line indicates the path used to extract the position-velocity diagrams shown in Figure~\ref{Pvdiag}. Along this curve, the intermediate endpoints are marked as A$_0$ and A$_1$. The beam size is indicated in the bottom-right of the map. The location of the massive cores within the cloud is indicated with light green crosses \citep{rathborne2006,butlerTan2012}. \textit{Right:} Velocity  map (color scale) of the H$^{13}$CO$^+$(1-0) emission with superimposed the dense gas tracer and mass surface density emission contours as in the left panel. Only pixels with integrated intensity above 3$\sigma$ are shown. The north and east directions are indicated in the bottom-right corner. In both panels, the SiO emission contours, from 3$\sigma$ ($\sigma$=0.07 K km s$^{-1}$) by steps of 3$\sigma$, and the 0.1 g cm$^{-2}$ mass surface density contour (visual extinction $\sim$20 mag) are superimposed \citep{kainulainen2013}.}
    \label{figH13COp}
\end{figure*}

From the SiO emission, we estimate the mass, momentum and energy of the SiO-traced gas toward G035.39-S, using the method described in \cite{dierickx2015}:

\begin{equation}
    M = \frac{d}{\chi(SiO)} \times \mu_g \ m(H_2) \times \Sigma_{pix} N(SiO)_{pix}
\end{equation}

\begin{equation}
    P = M\ \mathrm{v}
\end{equation}

\begin{equation}
    E = \frac{1}{2} M \ \mathrm{v}^2
\end{equation}

\noindent 
where d is the source kinematic distance \citep[2.9 kpc;][]{simon2006}, $\mu_g$=1.36 is the gas molecular weight, m(H$_2$) is the molecular hydrogen mass, $\chi$(SiO) is the SiO fractional abundance with respect to H$_2$ and $\Sigma_{pix} N(SiO)_{pix}$=(3.5$\pm$0.7)$\times$10$^{13}$ cm$^{-2}$ is the total SiO column density, summed for all pixels above 3$\sigma$. Finally, v=8 km s$^{-1}$ is the linewidth at the base (3$\times$rms) of the emission in the SiO spectrum averaged across G035.39-S. Similarly to what described in \cite{cosentino2022}, we have estimated $\chi$(SiO) as follows:

\begin{equation}
    \chi(SiO) = \frac{N(SiO)}{N(^{13}CO)} \times \chi(CO) \times \frac{^{13}C}{^{12}C}
\end{equation}

\noindent
where N(SiO)=(1.9$\pm$0.2)$\times$10$^{12}$ cm$^{-2}$ and N(H$^{13}$CO$^+$)=(2.0$\pm$0.3)$\times$10$^{12}$ cm$^{-2}$ have been calculated toward the SiO emission peak, using equation A4 in \cite{caselli2002}. Both this and the previous values of SiO column densities have been calculated assuming Local Thermodynamic Equilibrium (LTE). We note that, since the SiO emission is due to shocks, LTE conditions may not be representative of the gas. However, a non-LTE analysis, i.e. a Large Velocity Gradient analysis with RADEX \citep{tak2007} would require observations of multiple SiO rotational transitions that are not available at this stage. Moreover, higher-J transitions such as the SiO(3-2) have been seen to be either very week or undetected toward G035.39-N \citep{jimenezserra2010}. This would most likely be the case also toward G035.39-S, where the SiO emission is even narrower. Observations of the SiO(1-0) transition would therefore be the most suitable but are not currently available. In addition to this, we also note that while the method described by \cite{dierickx2015} is typically used to infer physical parameters in molecular outflows, this has also been used in the literature for other shell-driven shocks and ultra-high velocity gas nearby SNRs \citep[e.g. ][]{Yamada2017,Tsujimoto2018}. Other methods may require a more in depth knowledge of the shock driving source, which is not the case for this work. Hence, we consider the estimates here provided as approximate estimates and the calculation will be revisited when more data will become available.\\ For the column density estimate, we assume excitation temperature T$_{\mathrm{ex}}$= 9 K, as reported by \cite{jimenezserra2010} for the narrow SiO emission toward G035.39.39-N. The assumed value is also consistent with the excitation temperature inferred by \cite{cosentino2018,cosentino2022} toward two low-velocity parsec-scale shocks driven by the SNRs W44 and IC443. In accordance with these studies, we also assume an uncertainty of 20$\%$ on T$_{\mathrm{ex}}$. The uncertainties on the column densities have been estimated by propagating this and an additional 10$\%$ uncertainty due to the rms. Finally, we assume $^{13}$C/$^{12}$C=53$\pm10$ \citep{milam2005} and $\chi$(HCO$^+$)=(1$\pm$0.5)$\times$10$^{-8}$ \citep{vanDishoeck1993}. With this method, we estimate an SiO fractional abundance $\chi$(SiO)=(1$\pm$0.6)$\times$10$^{-10}$, which in turn gives mass, momentum and energy of the SiO-traced gas M=175$\pm$90 M$_{\odot}$, P = (1.4$\pm$0.8)$\times$10$^3$ M$_{\odot}$ km s$^{-1}$ and E=(1.1$\pm$0.6)$\times$10$^{41}$ erg. We note that this energy estimate is the kinetic energy of the SiO-traced gas and does not correspond to the full energy released by the shock. Indeed, only a small fraction of the shock energy is expected to be dissipated through gas turbulent motion \citep{Park2019}. Finally, for our estimates of P and E, we have considered the width of the SiO(2-1) line at the base of the emission. If we were to consider only the energy contribution from the SiO gas bulk motion then we would need to consider as typical linewidth the average SiO velocity dispersion across the cloud $\sigma_\mathrm{v}\sim$1.5-2 km s$^{-1}$ and the energy would be E=3/2 M$\sigma_\mathrm{v}$. We would then obtain P$\sim$400 M$_{\odot}$ km s$^{-1}$ and E$\sim$10$^{40}$ erg. These values are also incompatible with those measured toward molecular outflows.\\

\noindent
Toward G035.39, \cite{NguyenLuong2011} investigated the 70 $\mu$m emission obtained with Herschel and reported the presence of 5 deeply-embedded protostars toward the south of the cloud, that the authors classified as IR-quiet massive dense cores. The energy of molecular outflows powered by high-mass \citep[$\sim$10$^{46}$ ergs;][]{zhang2005, lopezsepulcre2009,Liu2025} and even intermediate-mass \citep[$\sim$10$^{43}$-10$^{44}$ ergs;][]{beltran2006,beltran2008} sources is several order of magnitude higher that the energy measured for the SiO-traced gas toward G035.39-S. Similarly, the mass and momentum here reported are up to 3 orders of magnitude higher than those typically reported for molecular outflows powered by low-mass protostars \citep{dunham2016}. We also note that no significant SiO(5-4) emission is present toward the cores in G035.39-S, as reported by \cite{liuMeng2018} using ALMA single-pointing observations. Hence, although some minimal contribution from molecular outflows may be present, it is unlikely that the SiO emission toward G035.39-S is dominated by ongoing star-formation activity. Therefore additional mechanisms need to be considered. 

\subsection{Signature of cloud-cloud collision in G035.39}
\begin{figure}[!htpb]
    \centering
    \includegraphics[width=\linewidth, trim= 5cm 0cm 3cm 0cm, clip=True]{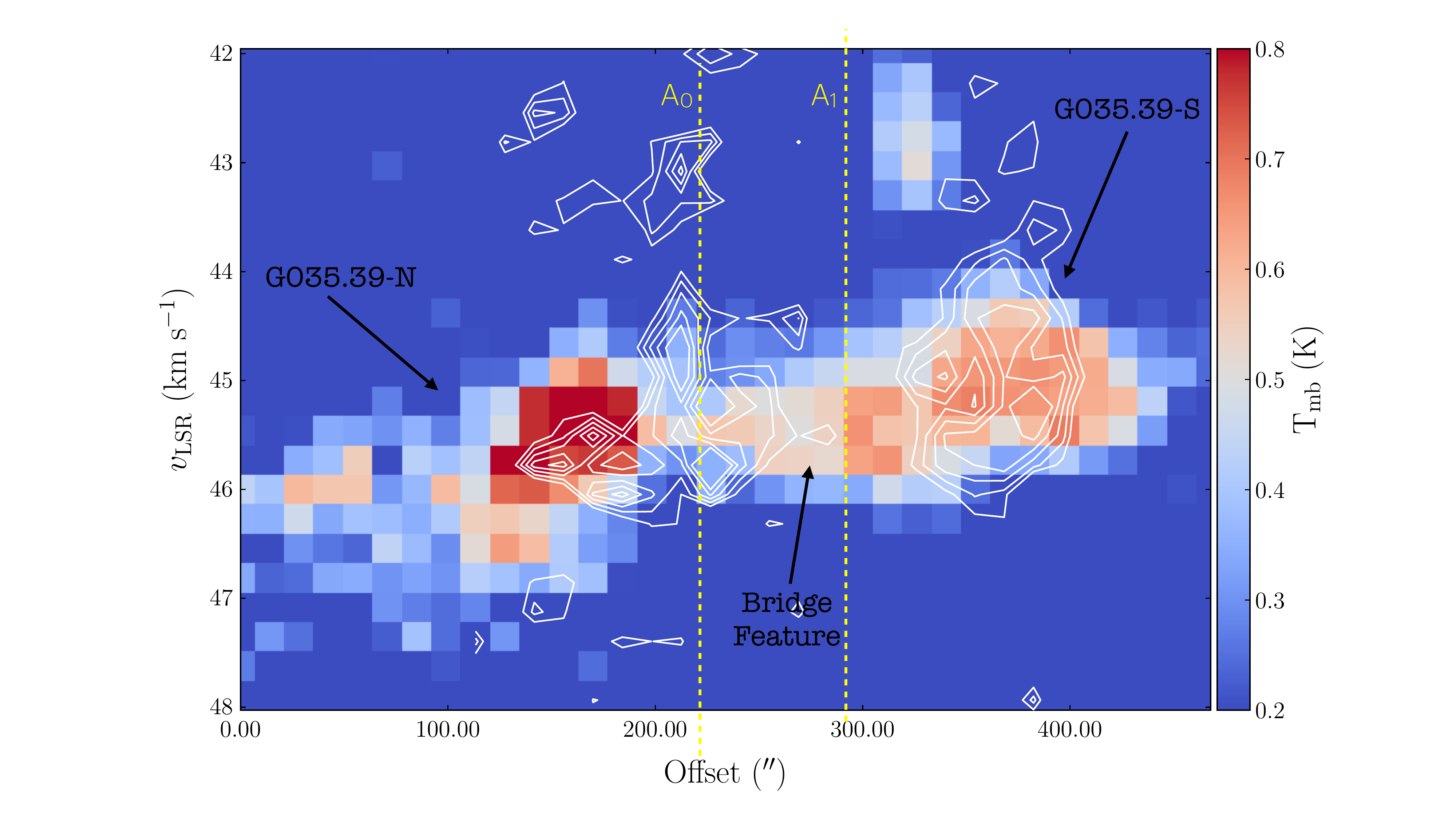}
    \caption{Position-Velocity (pv) diagram of the H$^{13}$CO$^+$ emission (color scale) extracted across the yellow curve seen in the left panel of Figure~\ref{figH13COp}. The verical dashed yellow lines correspond to the endpoints A$_0$ and A$_1$ indicated in Figure~\ref{figH13COp}. The white contours corresponds to the SiO pv diagram extracted along the same curve from 3$\sigma$ ($\sigma$=0.03 K) by steps of 3$\sigma$.}
    \label{Pvdiag}
\end{figure}

We now study the morphology and kinematics of the dense gas tracer H$^{13}$CO$^+$(1-0) across the full extent of G035.39. In Figure~\ref{figH13COp}, we show the spatial distribution (integrated intensity map over the range 42-48 km s$^{-1}$; left panel) and velocity field (moment 1 map; right panel) of the dense gas tracer emission. In both panels, the SiO emission contours (black) from Figure~\ref{figSiO} and the 0.1 g cm$^{-2}$ mass surface density contour (cyan) are superimposed. From Figure~\ref{figH13COp}, the dense gas tracer emission is organised into two filamentary structures that follow well the G035.39-N and G035.39-S morphologies, respectively. Again, while the northern filament appears almost vertical across the north-south direction, the southern filament is tilted toward the east by $\sim$45$^{\circ}$ with the respect to it. As seen from the H$^{13}$CO$^+$ velocity map, the G035.39-N and G035.39-S filaments do not just appear as having different spatial orientation but they also are kinematically separated by $\sim$1-1.5 km s$^{-1}$, with velocities of $\sim$45.5-46.5 km s$^{-1}$ and $\sim$44.5-45 km s$^{-1}$, respectively. This velocity separation is $>$3$\delta$v and hence significant within the velocity resolution. Toward the centre of the source, the two components mix and the dense gas presents intermediate velocities between $\sim$45-45.5 km s$^{-1}$. This kinematic structure is similar to that expected in cloud-cloud collisions events, i.e., two velocity components connected by a fainter bridge emission. Hence, we have extracted a position-velocity (pv) diagram across the yellow curve in Figure~\ref{figH13COp}, and this is shown in Figure~\ref{Pvdiag}. 

\noindent
From the pv diagram in Figure~\ref{Pvdiag}, the bridge emission connecting the northern and southern filaments of G035.39 is clearly seen. The narrowest SiO line profile detected across the cloud is spatially associated with the bridge feature. This region was not covered by the map presented in \cite{jimenezserra2010} and it is at the very edge of the SiO ALMA map that will be presented by Liu et al. in prep. On the other hand, the SiO emission toward G035.39-S does not seem either spatially nor kinematically associated with the bridge feature. As already reported in previous sections, this emission is not due to ongoing star-formation activity and it shows a different orientation with respect to the northern emission. Moreover the SiO emission toward G035.39-S is on over-all slightly blue-shifted compared to the bulk of the northern SiO emission (see spectra A, B, C versus spectra D, E an F in Figure~\ref{figSiO}). Hence, although the two emissions are not completely kinematically separated, it may be that the bulk of the northern and southern SiO emission have different drivers.
 
\subsection{The G035 complex: bubble-driven cloud-cloud collision?}\label{index}
\begin{figure}[!htpb]
    \centering
    \includegraphics[width=\linewidth,trim= 9cm 0.5cm 3.5cm 0cm, clip=True]{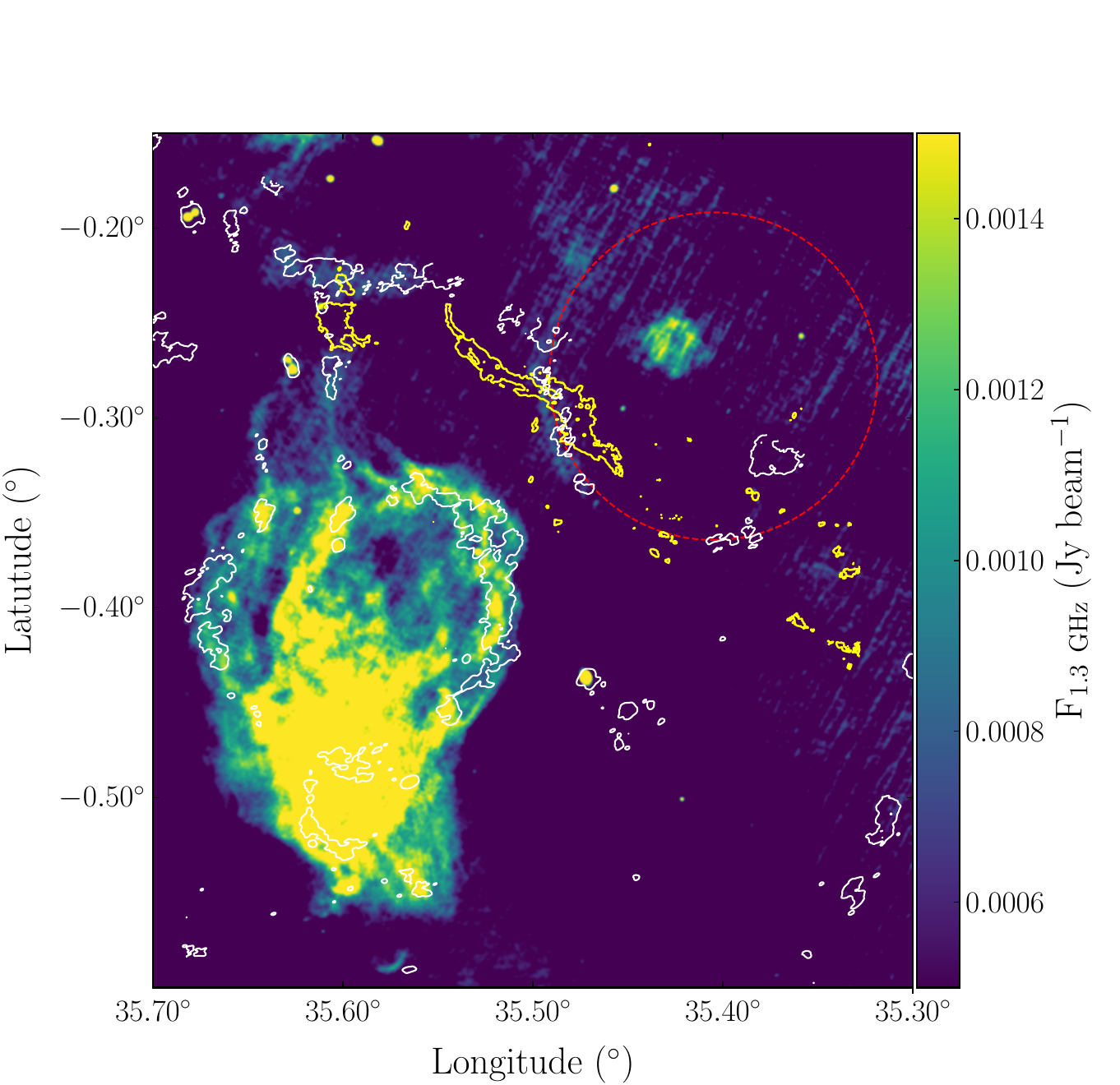}
    \caption{1.3 GHz emission (color scale) image obtained with MeerKat \citep{Goedhart2024} toward G035.39 and its surroundings. The yellow contours corresponds to 0.1 g cm$^{-2}$ in the mass surface density map of \cite{kainulainen2013} and highlights the shape and location of the IRDC. The white contour corresponds to the 3$\sigma$ ($\sigma$=0.0004 ) emission level in the Giant Metrewave Radio Telescope (GMRT) image at 610 MHz from \cite{Paredes2014}. The red dashed circle indicates the location of the putative bubble.}
    \label{MeerkatLarge}
\end{figure}

In order to identify these shock drivers, we have inspected the publicly available MeerKat 1.3 GHz image \cite{Goedhart2024} toward and around G035.39, as shown in Figure~\ref{MeerkatLarge}. 

\noindent
As seen from Figure~\ref{MeerkatLarge}, G035.39 is located within a complex network of Galactic bubbles \citep{Shen2024}. The brightest of these, located toward south-east of the image, is the SNR G035.6-0.4 \citep{Paredes2014,green2019}. The compact source toward north-west is the H{\small II} region G035.43-00.26, already reported in the WISE Catalogue of of H{\small II} regions by \cite{anderson2014}. Toward the same region, a fainter bubble-like structure (red dashed circle) is present both in the MeerKat and GMRT images (see color scale and white contours respectively), spatially associated with G035.39-S. This structure is not reported neither in the WISE Catalogue of H{\small II} regions \citep{anderson2014} nor in the Catalogue of Galactic Supernovae by \cite{green2019} and most likely it has not been classified yet. We note that this faint bubble is not associated with the compact, bright source also encompassed by the red dashed circle. This additional compact source is indeed the H{\small II} regions G034.4+0.23, which has V$_{LSR}\sim$60 km s$^{-1}$ \citep{watson2003}, i.e. it is kinematically distinct from both the gas associated with G035.39 and the fainter bubble. A rough estimate of the spectral index at different locations across the source gives $\alpha\sim$-0.5{\footnote{The spectral index has been estimated as $\alpha = \mathrm{log(F_1/F_2)/log(\nu_1/\nu_2)}$. Where F$_1$, $\nu_1$ and F$_2$, $\nu_2$ are the flux in Jy and frequency in MHz of the Meerkat and GMRT data, respectively.}}, consistent with typical values measured in SNRs \citep{Ranasinghe2023}. Further investigations are needed to fully characterise the nature of the bubble. \\

\begin{figure*}[!htpb]
    \centering
    \includegraphics[width=0.5\linewidth,trim= 2cm 0cm 3.5cm 0cm, clip=True]{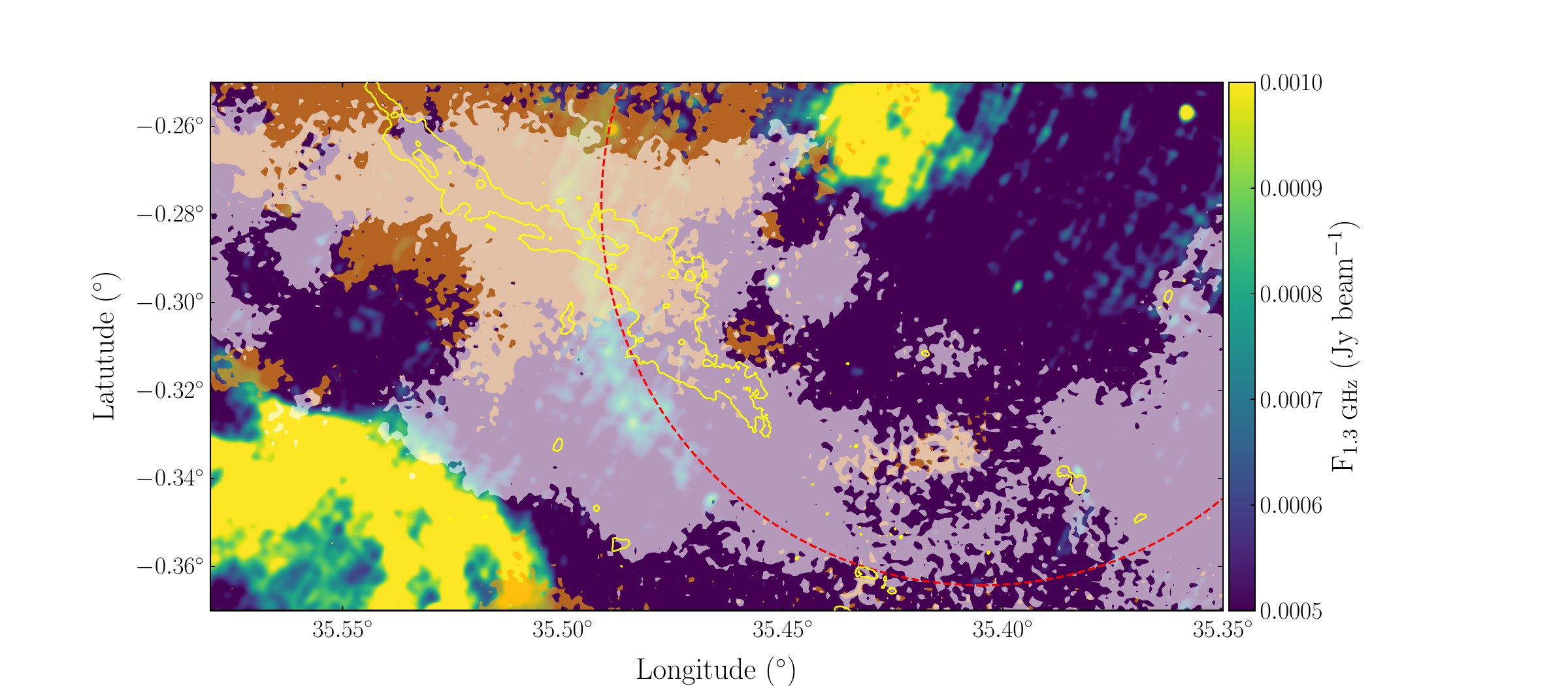}\includegraphics[width=0.5\linewidth,trim= 2cm 0cm 3.5cm 0cm, clip=True]{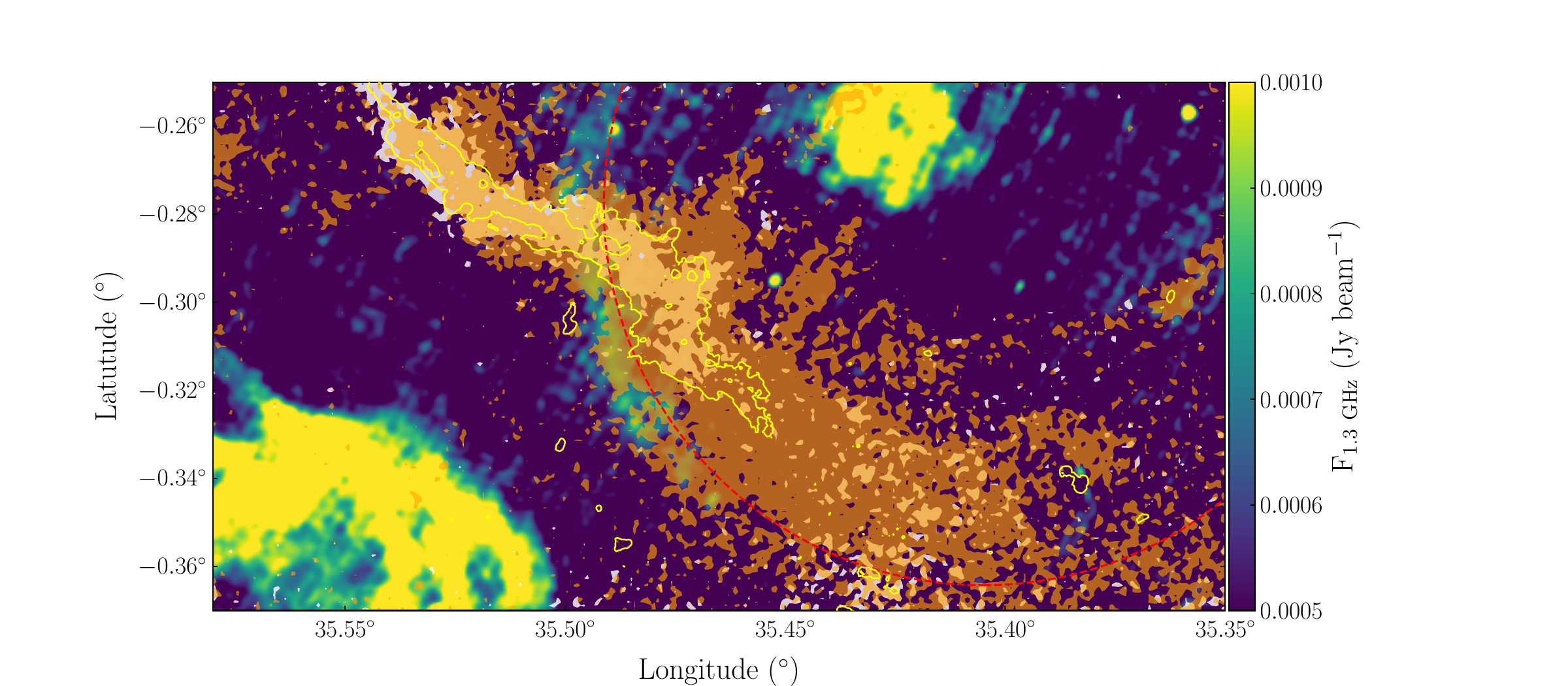}
    \caption{1.3 GHz emission (color scale) image obtained with MeerKat \citep{Goedhart2024} toward G035.39. The yellow contours corresponds to 0.1 g cm$^{-2}$ in the mass surface density map of \cite{kainulainen2013} and highlights the shape and location of the IRDC. The white (44-45.5 km s$^{-1}$) and orange (45.5-46.5 km s$^{-1}$) shadows correspond to the integrated intensity maps of the $^{13}$CO(1-0) (left panel) and  C$^{18}$O(1-0) emission (right panel) obtained with the Green Bank Telescope 100 m antenna. The orange shadows correspond to integrated emission larger than 3$\sigma$  for both the $^{13}$CO ($\sigma$=0.4 K km s$^{-1}$) and C$^{18}$O emission ($\sigma$= 0.15 km s$^{-1}$). The white shadows correspond to integrated emission larger than 6$\sigma$ for  the $^{13}$CO emission ($\sigma$=0.5 K km s$^{-1}$) and 3$\sigma$ for the C$^{18}$O emission ($\sigma$= 0.17 km s$^{-1}$). }
    \label{MeerkatSmall}
\end{figure*}

\noindent
In Figure~\ref{MeerkatSmall}, a zoom-in view of the MeerKat image, centred on G035.39, is displayed. Overlaid on this, we show integrated intensity maps of the lower-dense gas tracer $^{13}$CO(1-0) (left panel) and C$^{18}$O(1-0) (right panel) emission obtained with the 100 m antenna at the Green Bank Telescope. The white and orange shadows correspond to integration ranges of 45.5-46.5 km s$^{-1}$ and 44-45.5 km s$^{-1}$ respectively, i.e., consistent with the velocity of the dense gas toward G035.39-N and G035.39-S. The two velocity structures are seen in both tracers, but they appear more extended in $^{13}$CO. For the integrated intensity maps, we have calculated the integrated noise as A$_{\mathrm{rms}}$=rms$\times\delta$v$\times \sqrt{N_{ch}}$. In both cases, the velocity resolution is $\delta$v$\sim$0.19 km s$^{-1}$, while the rms is 0.3 K and 0.6 K for $^{13}$CO and C$^{18}$O, respectively. N$_{ch}$ corresponds to the number of channels within each integration range, i.e. N$_{ch}$=13 and 6 for the map integrated over the velocity range 45.5-46.5 km s$^{-1}$ and N$_{ch}$=24 and 8 for the map obtained over the velocity range 44-45.5 km s$^{-1}$, for $^{13}$CO and C$^{18}$O respectively. From Figure~\ref{MeerkatSmall}, the C$^{18}$O emission toward G035.39-S (orange shadow) follows well the curve of the shell seen at 1.3 GHz (red circle), as well as the lower mass surface density material of the cloud ($\Sigma <$0.1 g cm$^{-2}$). Hence, the identified bubble, not just spatially overlaps with the G035.39-S, but they also seem to be kinematically associated. Therefore, the dense filament seen in H$^{13}$CO$^+$ toward G035.39-S is most likely the densest part of a much larger filament that is being pushed by the expansion of the bubble.\\

\begin{figure}
    \centering
    \includegraphics[width=\linewidth,trim= 12cm 0.5cm 10cm 0cm, clip=True]{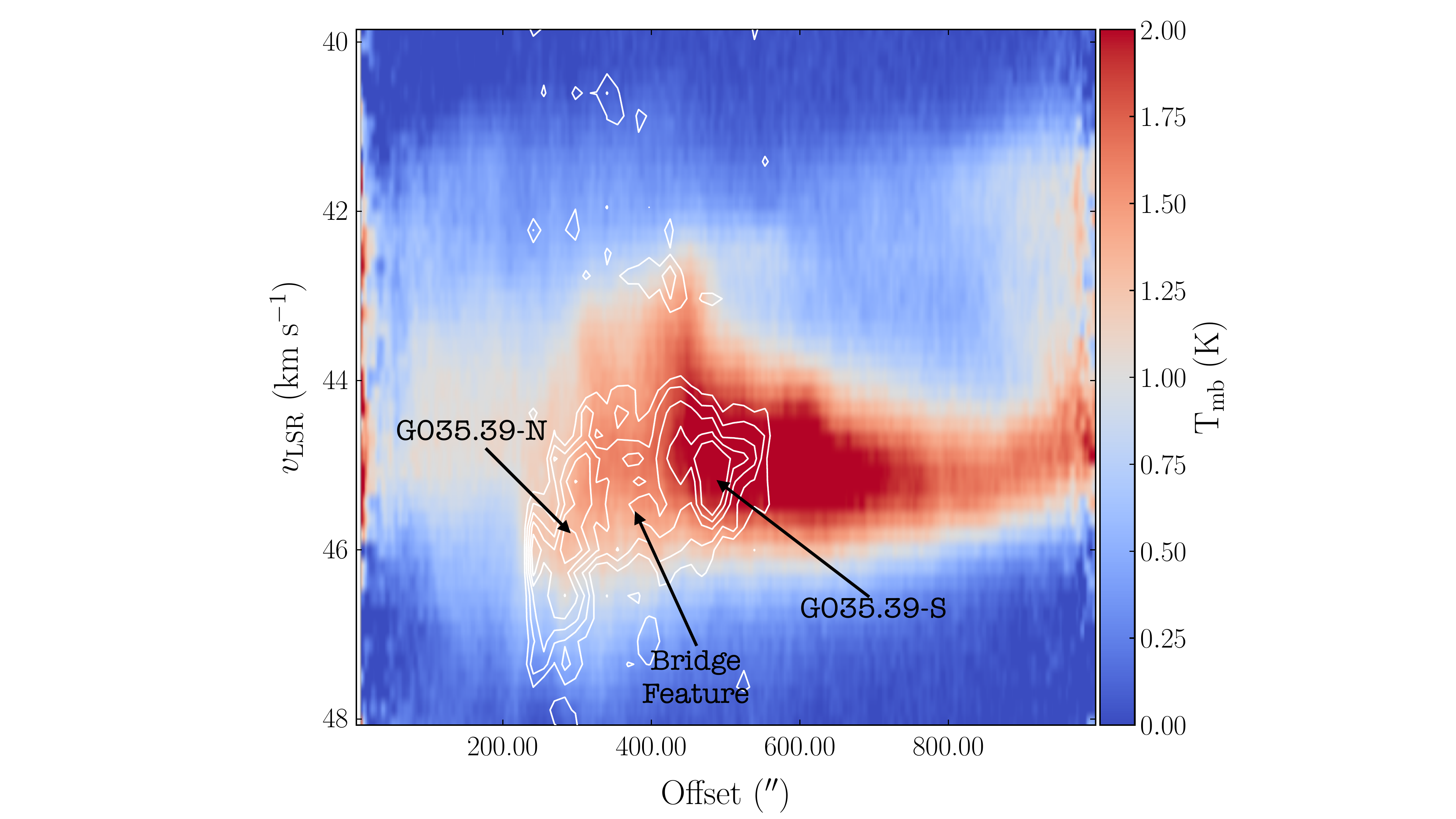}
    \caption{Pv-diagram obtained by collapsing the $^{13}$CO(1-0) (color scale) and H$^{13}$CO$^+$(1-0) (white contours) cubes along the Declination axis. H$^{13}$CO$^+$ contours are from 3$\sigma$ ($\sigma$=0.015 K) by steps of 3$\sigma$.} 
    \label{pvCO}
\end{figure}

\noindent
Additional evidence of shock compression in the gas kinematics are usually seen in the form of line broadening and/or by the presence of cavity-like structure in position velocity diagrams \citep{Chen2017,Fukui2018b,Chen2025}. Due to the relatively narrow linewidth of the SiO-traced gas, this is likely driven by a low velocity shock. Hence, significant line broadening is not expected nor it is observed in the $^{13}$CO and C$^{18}$O(1-0) emission. In order to identify cavity-like structures, in Figure~\ref{pvCO} we show the pv-diagram obtained from the $^{13}$CO(1-0) cube along constant Declination (color scale). The same pv-diagram has been obtained for the H$^{13}$CO$^+$(1-0) emission (white contours).

\noindent
From Figure~\ref{pvCO}, the $^{13}$CO kinematic structure hints to the presence of two cavity-like structures are identified in the $^{13}$CO pv-diagram. An almost vertical cavity is in spatial contact with G035.39-N that corresponds to the direction in which the SNR G35.6-0.4 is expanding (Liu et al. in prep.). A horizontal, more extended, cavity-like structure is in spatial contact with G035.39-S and corresponds to the expanding shell of the putative bubble identified in this work. A more detailed analysis of the lower-density gas kinematics is needed to fully ascertain the link between the two cavity-like structures and the filaments. This analysis will be presented in a forthcoming work (Law et al. in prep.). The H$^{13}$CO$^+$ pv-diagram extract along the same direction is also shown in Figure~\ref{pvCO} (white contours). The two filaments, G035.39-N (red-shifted) and G035.39-S (blue-shifted), connected by the bridge-like feature are clearly identified. The overall H$^{13}$CO$^+$ kinematics resemble that of a V-shaped gas distribution, with the V pointing toward north. Such a kinematics has been reported as associated with cloud-cloud and filament-cloud collision events \citep{Arzoumanian2018,fukui2018}.

\section{Discussion and conclusions}\label{discussion}
The mechanisms responsible to initiate star formation in IRDCs are still unknown. Slow shocks extending over parsec-scales and impacting on molecular clouds may efficiently regulate star formation in these objects as well as shaping their morphology \citep[e.g.][]{Khullar2024}. Both cloud-cloud collision events and expanding shells of Galactic bubbles are regarded as drivers of such shocks \citep{inutsuka2015} and have been proposed as key ingredients to initiate star formation in molecular clouds. In this work, we have looked for signature of large-scale low-velocity shocks toward the IRDC G035.39-00.33. We have used IRAM-30m observations of the SiO and H$^{13}$CO$^+$ emission to investigate the morphology, kinematics and line profiles of the shocked and dense gas across the cloud. 

\subsection{The origin of the SiO emission in G035.39}
From the analysis of the dense gas emission, we find that the cloud is organised into two filaments, the northern (G035.9-N) and the southern (G035.39-S) filaments, tilted by $\sim$45$^{\circ}$ with respect to each other. Moreover, while G035.9-N has velocities between $\sim$45.5-46.6 km s$^{-1}$, the southern filament is slightly blue-shifted (45.5-44 km s$^{-1}$). Toward the centre of the cloud, the two filaments coexist and seem to be connected by a fainter bridge-like emission, also seen in the position velocity diagram. The narrow SiO emission identified across the cloud is spatially and kinematically coincident with this bridge emission. This is clear evidence of an ongoing cloud-cloud interaction toward the IRDC, i.e. the two filaments represent the densest part of two colliding clouds, also seen in C$^{18}$O(1-0) emission, while the narrow SiO emission traces the shock interface between them. This confirms the scenario of cloud-cloud collisions toward the IRDC G035.39 previously suggested by \cite{jimenezserra2010} and later by \cite{henshaw2014,bisbas2018,LiuTie2018} through indirect evidence.\\ Besides the very narrow SiO emission associated with the bridge-like feature and located toward the center of the cloud, we detect additional SiO-traced gas emission widespread across the cloud. The broad and narrow SiO emission toward G035.39-N was first identified by \cite{jimenezserra2010} and recently studied at high-angular resolution by Liu et al. in prep. Taking advantage of the high-angular resolution of the ALMA images, the authors confirmed that the broad SiO emission is powered by molecular outflows, while the narrow ($<$2 km s$^{-1}$) component is most likely due to the ongoing interaction between G035.39-N and the nearby SNR G035.6-0.4 \citep{Paredes2014,green2019}. Toward G035-39-S, we also find SiO emission whose line profiles are narrower that 3 km s$^{-1}$ and that is extended over more than 2 pc. Note that narrow line profiles with typical FWHM$\leq$3 km s$^{-1}$ have been found to be associated with the interaction between the SNR W44 and the IRDC G034.77-55 \citep[][see below]{cosentino2018,cosentino2019}. Also, 
the mass, momentum and energy estimated for the SiO-traced gas are not consistent with those typically observed in molecular outflows associated with low-, intermediate and high-mass forming stars. Hence, on-going star formation activity from the five deeply embedded sources identified is unlikely to be the only driver of the SiO-traced gas emission toward G034.39-S. A similar argument was presented by \cite{NguyenLuong2011} for the northern narrow SiO emission. Furthermore, the energy and momentum associated with the SiO emission are consistent with those observed toward other SNR-driven shocks \citep{sashida2013,cosentino2022}. However, with the data currently available it is not possible to completely rule out the star formation activity scenario. We cannot exclude the presence of additional, unresolved and deeply embedded low-mass protostars that may be driving outflows and hence the observed SiO emission, as it has been seen toward other IRDCs \citep[e.g.,][]{foster2014}. Future higher-resolution observations targeting both the molecular gas and embedded continuum sources, as well as sensitive mid-IR observations with JWST, will be crucial to disentangle between the two scenarios.

\subsection{Stellar-feedback-driven cloud-cloud collision}
The IRDC G035.39 is part of the more extended G035 complex and it is surrounded by both H{\small II} regions and SNRs. In particular, we find that the southern filament G035.39-S is spatially and kinematically associated with a shell-like structure not yet reported in the most up-to-date catalogue of galactic H{\small II} regions and SNRs \citep{anderson2014,green2019}. Our rough estimated of the bubble spectral index ($\alpha\sim$-0.5; Section~\ref{index}) seems to indicate that this may be the relic of a SN explosion. This is not unlikely since within the immediate surroundings of G035.39 at least two additional SNRs of similar age (20-30 kyr) are present i.e., G035.6-0.4 \citep{Paredes2014,green2019} and W44 \citep{wootten1978,cosentino2019}. The SNR W44 is interacting with the know IRDC G034.77-00.55 and, at the shock interface, the SiO linewidths are similar to those reported here toward G035.39 \citep{cosentino2018,cosentino2019}. We also note that the narrow SiO emission detected toward G035.39-S is consistent with that identified toward other IRDCs embedded into bubbles networks \citep{cosentino2020}. The mass, momentum and energy here estimated for the SiO-traced gas are also consistent with those reported by \cite{cosentino2022} toward the interaction site between the SNR IC443 and the clump G. We thus conclude that the SiO emission toward G035.39-S is likely due to the interaction between this putative SNR and the cloud. Hence, the IRDC G035.39 consists of two colliding filaments, pushed toward each other by nearby external stellar feedback in the form of SNRs, i.e., the SNR G035.6-0.4 pushing the northern filament and the putative unclassified bubble sweeping up the southern filament. This event may therefore be considered a stellar-feedback-driven cloud-cloud collision, i.e., not due to the clouds natural shear motion but rather to galactic expanding gas. This scenario seems to be also supported by the presence of cavity-like strcuture in the $^{13}$CO pv-diagram. However, the highly complex kinematic structure of the cloud \citep{LiuTie2018} needs to be investigated more in details and at higher-angular resolution. Therefore, at this stage we cannot fully exclude the alternative scenario that the monolithic, pre-existent G035.39 filament may have been impacted by the nearby expanding shell, causing the observed bent structure toward the south of the cloud.\\
\noindent
The idea that cloud-cloud collisions may occur as a consequence of the expansion of galactic shells into the interstellar medium was first introduced by \cite{inutsuka2015} and more recently observationally investigated by \cite{cosentino2020}. Toward G035.39, our results support the idea that a stellar-feedback-driven cloud-cloud collisions is occurring and that this may have helped to assemble the IRDC with its current morphology. Hence, expanding on what suggested in \cite{cosentino2020}, this type of cloud-cloud collisions may also need to be considered as an IRDC formation mechanisms and further theoretical and observational studies are needed. Observational evidence of dense clouds located in between multiple bubbles have already been reported, for instance toward the W41 complex \citep{Hogge2019} and W44 \citep{cosentino2019}, and they may not be rare. However, our study suggests that, while stellar-feedback-driven cloud-cloud collisions leave on the newly formed IRDC similar footprint than those detected in natural occurring cloud-cloud collisions, a multi-wavelengths approach is crucial to discern between the two scenarios.\\ Finally, we note that with the data in hand, it is challenging to asses whether the observed low-velocity, large-scale shocks may have ignited the ongoing star formation within G035.39. The location of the massive cores reported by \cite{NguyenLuong2011}, well follows the orientation of the SiO-traced gas emission toward G035.39-S. These sources are classified as IR-quiet, hence they may not have reached the proto-stellar phase yet. Toward G035.39-N, Liu et al. in prep. estimate the dynamical age of the molecular outflows to be lower than the SNR age. Hence, it is possible that some star formation may have been triggered by the stellar feedback cloud-cloud collision. However, higher-angular resolution images of dense gas tracers, such as N$_2$H$^+$, are needed to resolve the typical core spatial scales at which these star formation signatures are seen.

\noindent
In conclusions, we report evidence of a cloud-cloud collision toward the IRDC G035.39 that may have been driven by the expanding shells of two nearby SNRs. The two colliding clouds are still visible in the kinematics of the dense gas emission. Low-velocity, large-scale shocks are seen both at the interface of the collision as well as toward the site of interaction between the expanding shells and the swept-up filaments. Higher-angular resolutions and multi-wavelengths observations are needed to unveil whether star formation has been triggered by these shocks. Our study poses the accent on how stellar feedback cloud-cloud collisions may be more ubiquitous that so far suggested and that multi-wavelengths, multi-spatial scale observations need to be combined to fully distinguish between stellar feedback and naturally occurring cloud-cloud collisions. 

\begin{acknowledgements}
This work is based on observations carried out under project number 134-12 with the IRAM 30 m telescope. IRAM is supported by INSU/CNRS (France), MPG (Germany) and IGN (Spain). J.C.T. acknowledges support from
ERC project 788829–MSTAR. I.J.-S. acknowledges funding from grant No. PID2019-105552RB-C41 awarded by the Spanish Ministry of Science and Innovation/State Agency of Research MCIN/AEI/10.13039/501100011033. J.D.H.
gratefully acknowledges financial support from the Royal Society (University Research Fellowship; URF/R1/221620). R.F. acknowledges support from the grants Juan de la Cierva FJC2021-046802-I, PID2020-114461GB-I00 and
CEX2021-001131-S funded by MCIN/AEI/ 10.13039/501100011033 and by “European Union NextGenerationEU/PRTR”. S.V. acknowledges partial funding from the European Research Council (ERC) Advanced Grant MOPPEX 833460.
S.V. and J.C.T acknowledge the support from a Royal Society International Exchanges Scheme grant (IES/R3/170325).      
\end{acknowledgements}

%
%

\bibliographystyle{aa} 
\bibliography{aa.bib}
\end{document}